\documentclass[apjl]{emulateapj}
\usepackage{epstopdf}
\usepackage{natbib}
\usepackage{amsmath}
\usepackage{enumerate}
\usepackage{cancel}

\def\simpropto{\lower.2ex\hbox{$\; \buildrel \propto \over \sim \;$}}
\def\ltsim{\lower.5ex\hbox{$\; \buildrel < \over \sim \;$}}
\def\gtsim{\lower.5ex\hbox{$\; \buildrel > \over \sim \;$}}

\usepackage{color}

\def\bea{\begin{eqnarray}}
\def\eea{\end{eqnarray}}
\def\simpropto{\lower.2ex\hbox{$\; \buildrel \propto \over \sim \;$}}
\def\ltsim{\lower.5ex\hbox{$\; \buildrel < \over \sim \;$}}
\def\gtsim{\lower.5ex\hbox{$\; \buildrel > \over \sim \;$}}
\begin{document}
\def\mnras{Mon. Not. Roy. Astr. Soc.}
\def\ap{ApJ}
\def\apjl{ApJL}

\title{ Feedback by Massive Black Holes in Dwarf Galaxies}

\author{ Joseph Silk}
\affiliation{
Institut d'Astrophysique,  UMR 7095 CNRS, Universit\'e Pierre et Marie Curie, 98bis Blvd Arago, 75014 Paris, France\\
{\color{black}AIM-Paris-Saclay, CEA/DSM/IRFU, CNRS, Univ Paris 7, F-91191, Gif-sur-Yvette, France
}\\
Department of Physics and Astronomy, The Johns Hopkins University,
Homewood Campus, Baltimore MD 21218, USA\\
Beecroft Institute of Particle Astrophysics and Cosmology, Department of Physics, University of Oxford, Oxford OX1 3RH, UK
}

\today
\begin{abstract}
Could there be intermediate mass black holes  in essentially  all old dwarf galaxies?  {
\color{black}
I  argue that current observations of Active Galactic Nuclei in dwarfs allow such a radical hypothesis which  provides early feedback and potentially provides a unifying explanation for many if not  all of the apparent dwarf galaxy anomalies, such as the abundance, core-cusp, "too big to fail", ultra-faint and 
baryon-fraction issues. I describe the supporting arguments, which are largely circumstantial in nature, and discuss a number of tests. There is no strong motivation for modifying the nature of cold dark matter in order to explain any of the dwarf galaxy "problems".}

\end{abstract}

\maketitle

\section{Introduction}

There are two types of observed  astrophysical black holes, stellar mass and supermassive.
Stellar mass black holes that originate in the deaths of stars in the $\sim 20-100\rm M_\odot$ range are observed directly via gravitational waves. Supermassive black holes (SMBH) are observed via kinematic signatures at the centers of galaxies, with most accuracy for the MWG and for nearby disks with orbiting circumnuclear masers. However broad emission line signatures provide compelling evidence for SMBH to high redshift in the cores of massive galaxies. The observed mass range of SMBH is $\sim 10^6-10^{10}\rm M_\odot.$ 

Logically, there is no reason not to expect a population of intermediate mass black holes  in the mass range $\sim 10^2-10^5\rm M_\odot$ that are too massive to be formed directly by stellar collapse in the recent universe. Indeed,
the formation of SMBH is believed to be  a combination of dynamical merging of smaller black holes and gas accretion.  The latter process is observed via x-ray and optical emission line studies. 

Gravity wave experiments, most notably LISA, will enable us to directly observe IMBH mergers.  The gas accretion contribution to the SMBH masses is constrained by the Soltan radiative argument \citep{soltan82}. Seed black holes of mass$\sim 10^3-10^4 \rm M_\odot$ are  usually required at high redshift in order to account for the existence and growth via accretion of SMBH in the mass range 
$\sim 10^9-10^{10}\rm M_\odot$  at $z\gtsim 6,$ unless super-Eddington growth is invoked.

IMBH are observed in dwarf galaxies, although the occupation fraction in uncertain. Early-forming  dwarfs are generally the building blocks of massive galaxies. Were they to universally contain IMBH, this would provide a reservoir for seeding growth of SMBH by tidal disruption of stars and gas accretion, as well as possibly by IMBH mergers. \textcolor{black}{
It is expected that many IMBH are left behind in the course of hierarchical merging 
 \textcolor{black}{\citep{rashkov2014} and should be present in many satellite dwarfs at $z=0$ \citep{wassenhove2010}}
as well as be  detectable as outliers on the black-hole scaling relations, especially at high redshift \citep{volonteri2009}.}

The best evidence for IMBH in dwarf galaxies comes from x-ray \textcolor{black}{ \citep{kormendyho2013, pardo16} }as well as optical  emission line studies. Active galactic nuclei are found in about 1\% of dwarfs, with the x-ray luminosities well in excess of  an x-ray  binary contribution as predicted from the star formation rate  \citep{baldassare16}. 
For any reasonable duty cycle, there must be IMBH in at least 10\% of dwarfs, and one cannot exclude a larger fraction because of the special circumstances required for the shallow gravitational potentials of dwarfs to retain gas as effectively as their massive counterparts. \textcolor{black}{ Moreover  a NUSTAR survey suggests that because of heavy x-ray absorption, a  significant number of dwarfs with AGN may be missed both in optical spectroscopic surveys and in 
X-ray surveys below 10 keV \citep{chen2017}.
Of course theoretical studies show that  not all dwarfs necessarily host IMBH \citep{volonteri2008}.}

Near IR studies of nearby dwarfs also suggest the presence  of obscured IMBH, but the data remains ambiguous \citep{hainlineA16}.
Optical studies of emission lines from central nuclei find a similar fraction  of IMBH in dwarfs to that found in x-ray surveys \citep{moran14}.

An interesting indirect argument for AGN feedback in dwarfs comes from the possibility of 
IMBH quenching of star formation in  dwarfs. The handful of identified dwarf IMBH lie on the usual $M_{BH}-\sigma$ scaling relation, where $M_{BH}$ is the black hole mass and $\sigma$ is the dwarf  stellar velocity dispersion.  However these same dwarfs fall below the $M_{BH}-M_\ast$ scaling relation where $M_\ast$ is the stellar mass {\color{black} \citep{baldassare15, reines15}. } One can  possibly interpret this as IMBH-induced suppression of star formation in dwarfs that occurred in the early gas-rich phase of dwarf evolution. 

IMBH are generically predicted to form in the course of Population III evolution. There is competition with fragmentation into Population III stars. Fragmentation dominates in the first generation of clouds unless there are special circumstances in which $H_2$ cooling is suppressed. However, clouds of mass $\sim 10^7-10^{8}\rm M_\odot$ that form at $z\sim 10$ collapse at virial temperatures  $\sim 10^4\rm K$ where $H_2$ formation indeed is suppressed.
IMBH plausibly form in these clouds, as fragmentation is  inhibited \citep{spaans06}. 
These systems are the precursors of many of  today's dwarf galaxies, which with measured or inferred halo masses of $\sim 10^8-10^{9}\rm M_\odot$ must have formed with baryon masses in the $\sim 10^7-10^{8}\rm M_\odot$ range.

Unless the clouds are contaminated by metal-injection  from  the smaller clouds at a level above $Z\sim 10^{-3}Z_\odot$, a process such as Lyman alpha photon  trapping and associated $H^-$photo-detachment \citep{johnson16} will guarantee high central cooling and accretion rates  leading to IMBH formation.  However the efficiency of  the IMBH production process is highly uncertain, particularly with regard to heavy element production by the first stars, followed by diffusion and mixing, as in  \cite {ferrara14}.  

In what follows, I will assume that the IMBH occupation fraction in dwarf galaxies is anywhere between 10\%  and 100\%.
There are a number of enigmas surrounding dwarf galaxies. Most  but not all of these have relatively conventional explanations in terms of baryonic feedback.  However in several cases, the modeling seems fine-tuned, and this has motivated many discussions of  modified dark matter as an alternative explanation. These modifications include appeals to warm \citep{bose17}, fuzzy \citep{marsh14} and self-interacting dark matter \citep{elbert15}, to name the most popular.

 In this note, I will reconsider baryonic feedback in the light of the possible presence of IMBHs.  First, I discuss the theoretical case for IMBH feedback in dwarfs. I then make the case that the presence of IMBH potentially provides a unifying explanation for essentially all of the dwarf galaxy anomalies. I conclude with several predictions that may help underpin and clarify the IMBH-dwarf  galaxy connection.

\section{The case for IMBH feedback in dwarfs}
A major question  concerning the efficiency  of AGN feedback is the efficiency of coupling relativistic jets and ultrafast nuclear outflows with the  ambient interstellar medium.
An example of efficient coupling is given by jet-driven backflows that lose enough angular momentum to feed  and self-regulate the central AGN \citep{cielo14}.
The efficiency problem may be   ameliorated in dwarfs which form at high redshift with high central density.
Outflows are more readily trapped and this  seems likely to lead to more coherent coupling  of the central source with the ambient medium.

These questions need to be explored in more detail. Here I will assume similar AGN jet or ouflow coupling to that normally adopted for massive galaxies, and only comment on the gross energetics, comparing supernova and AGN-driven feedback in gas-rich dwarf galaxies.

Consider a cloud of mass $M_c/10^8 \rm M_\odot$. 
This forms stars at $\sim 1\%$  efficiency and produces
$10^4$ SNe in a star cluster of $10^6\rm M_\odot$. These supernovae  generate $10^{55}$ ergs.

Compare this to  the IMBH case, where the energy produced by the AGN is   $\eta M_{BH}c^2 $ or $10^{56}M_3 $ ergs. Here the energy ejection efficiency $\eta=0.1$  and $M_3=M_{BH}/10^3\rm M_\odot.$  

Now consider superbubbles, which allow for the coherent explosions of supernovae, thereby enhancing their feedback efficiency into the surrounding galaxy.
The luminosity in mechanical energy production due to supernovae in a massive young star cluster is 
$$L_{SN} = 10^{40} \rm ergs/s (E_{SN}/10^{51} ergs)(\epsilon_{sfr}/0.01)(M_c/10^8\rm M_\odot) $$ 
over $40.10^6$ yr, the typical lifetime of an OB association.

An AGN also produces a superbubble. Its luminosity is 
$$L_{AGN}= (v/c)L_{edd=} 10^{42} {\rm ergs/s }(M_{BH/}10^5{ \rm M_\odot})(v/0.1c).$$
{\color{black} Even at an Eddington efficiency of 10\%, the }AGN    power gain amounts to  an order of magnitude  over that of supernovae. This will be aggravated by the fact that the denser interstellar medium in young dwarf galaxies will generate enhanced cooling and the efficiency will be correspondingly reduced, {\color{black} as well as possibly limiting central BH growth \citep{prieto17}.} The opposite effect is expected for AGN where the denser ISM reduces the porosity of the shocked medium. These effects need to be modeled in detail.

\section{ The circumstantial evidence for ubiquitous IMBH in the centers of dwarfs}

Theoretical reasoning aside, the main point I wish to make is that there is strong empirical but circumstantial evidence for AGN feedback in dwarfs. I review 10 observations of  problems which  confront the standard 
$\Lambda CDM$ model of structure formation  on dwarf galaxy scales. None of these can be said to be rigorously resolved. However most can be accounted for by supernova-driven feedback in combination with other plausible environmental processes such as ram pressure and tidal stripping. But at the very least, considerable fine-tuning is required. Some problems, previously thought to be  understood,   even resurface as simulation resolution, interstellar physics and observational diagnostics are improved.

The net effect is that a large community has seized on the notion that additional physics is needed that can only come from modifying the nature of dark matter in a drastic way. 
However modifying the nature of dark matter should be the last resort. Bringing in  fundamental physics to  solve astrophysical problems demands cast-iron evidence.  Einstein struck gold, because the advance of Mercury's precession fell into this category. However the dwarf galaxy evidence is far removed from such a degree of robustness.

In fact, I will show here that  a new astrophysical ingredient, IMBH feedback, provides a unifying theme that  can account for most, if not all, of the dwarf galaxy issues. No single example is robust, since  most of the details have yet to be explored. However I will argue that the circumstantial evidence is overwhelming for some 10 key dwarf galaxy issues that may be explained  in terms of a single unifying hypothesis. I will then present several potential tests of this hypothesis.

I begin by remarking that IMBH, which I postulate to be ubiquitously present in essentially all early-forming dwarfs,  are mostly passive today but were active in their gas-rich past.

\begin{enumerate}
\item
Feedback from IMBH can suppress the excess number of luminous dwarfs. The details have not been thoroughly studied, but by comparison with the controversial role of supernovae, an additional source of feedback provided by IMBH may be welcome. Simulators indeed  differ on whether supernova feedback is  effective: Studies such as 
\cite{oman16} argue positively, whereas \cite{bland15} and \cite{trujillo16} take the opposite view for the case of a more realistic multiphase ISM. The situation is equally unclear even  for massive disk galaxies in so  far as whether SNe are effective in driving gas fountains from the disk vis superbubbles, but compare  \cite{ostriker16} and \cite{keller16}, the latter arguing for the need for additional AGN feedback.
\item
 Discoveries of  ultrafaint dwarfs  in our galactic  halo and that of M31 are usually attributed to the success of supernova feedback. The recent discovery of  ultradiffuse massive dwarf galaxies poses more of a challenge \citep{vandokkum15}. 
 The feedback has clearly been dramatic in the past, and seems beyond the reach of supernovae, as  is  also the case for the "too-big-to-fail" problem, see below. A closely related phenomenon is the prevalence, at the $\sim 30\%$ level,  of  bulgeless disks often found  outside rich clusters \citep{kormendy16}.
These have formed their stellar disks  more efficiently, presumably via late infall of gas, 
 but early  bulge formation, long thought to be generic in CDM modelling of galaxy formation, must have somehow  been avoided. Again, effective SMBH or IMBH feedback is a plausible culprit, assuming that IMBH formation precedes the bulk of star formation.
 
 \item
  Numerical simulations predict more massive dwarfs than are observed, the so-called  Òtoo-big-to-failÓ problem \citep{boylan11}, whereby SNe are commonly believed to be  incapable of inhibiting star formation in  massive dwarfs \citep{garrison13}.
One  preferred solution is to modify dwarf kinematics by  SNe physics, as below,  in combination with  environmental evolution, notably tidal and ram pressure stripping \citep{brooks14}, although this begs the question of what happens in less dense environments such as the Local Group \citep{garrison14}.
 
\item
Supernova momentum input drives bulk gas motions in gas-rich dwarfs. The resulting dynamical heating of the dark matter generates cores in dwarfs \citep{pontzen14}.

In fact, AGN are equally capable of  
driving bulk gas motions \citep{peirani08}.  \textcolor{black}{Episodes of such motions dynamically heat the dark matter.
This occurs either via mechanically-driven  winds or by jet-driven bow shocks via gas cloud interactions, as observed  in  nearby radio galaxies such as 3C293 \citep{mahony16} and IC5063 \citep{morganti15}. }Modified dark matter in the form of WDM fails to generate large enough cores, but both fuzzy and self-interacting dark matter have been advocated as alternative solutions of the cusp/core problem \citep{schneider16}.

\item
A fraction $\sim$30 percent of baryons is  missing within the virial radius of massive disk galaxies \citep{bregman15}.
It is non-trivial to reduce baryon fraction in Milky Way-type galaxies, that is galaxies of Type Sb or later.  SNe fail to eject enough baryons: at best, they drive a fountain since the central SMBH is too small to induce  strong global feedback. IMBH feedback in gas-rich dwarfs provides an attractive solution, as the AGN  operate in the hierarchical assembly phase of disk galaxies
\citep{peirani12}.

\item
Identification of the sources responsible for the reionisation of  the universe remains a problem,
despite the reduced optical depth recently measured by the Planck satellite  \citep{planck16}.
While there are adequate numbers of galaxies present at the relevant epoch, $z \sim 8-10$, one has no idea of the appropriate escape fraction of ionizing photons. The problem of the unknown escape fraction for high redshift dwarfs is compounded by the near universal detection of Lyman-alpha halos by MUSE observations  of galaxies at intermediate redshift \citep{wisotzki16}, suggestive of a  low escape fraction of ionizing photons. This conclusion is however sensitive to the unknown porosity of the ISM. {\color{black} AGN (in dwarfs) } provide a potentially important source of ionizing photons, energetically capable of escaping the host galaxies {\color{black} \citep{volonteri09, madau15}.}

\item
Early chemical evolution generally occurs by bursts of star formation and accompanying supernovae  in dwarf galaxies that disperse much of the enriched debris into successive generations of subhalos and halos. The extreme low surface brightness MWG dwarf Reticulum II   with only $\sim$1000 stars presents an interesting exception, being  self-enriched in 
r-process elements such as europium \citep{ji16}. 

Reticulum II must have retained the enriched debris from an early binary neutron star merger. Such retention requires both a  dense ISM and an initially deeper potential well. Here one can imagine that early SNe must have failed to cause significant gas ejection, reminiscent of the "too-big-to-fail" problem. The  intervention of an AGN  might have plausibly provided the new ingredient that heated the dark matter potential well and triggered delayed escape both of gas directly and of stars via tidal disruption.

\item
Formation of SMBH at high redshift, {\color{black} $z\gtsim 6,$  }in the mass range $\gtsim  10^9 \rm M_\odot $, may require the presence of seed IMBH {\color{black} \citep{johnson2013}}.
The only alternative is super-Eddington accretion.

\item
AGN triggering of star formation is occasionally invoked to explain unusually efficient episodes of star formation, especially when there is morphological evidence of causality for  positive feedback induced by jets or fast outflows.   Examples include Minkowski's object, 3C 285 \citep{salome15}, Centaurus A \citep{crockett12}, 4C41.17 \citep{bicknell00} and  NCG 5643 \citep{cresci15}.

Positive feedback could in principle occur in dwarfs  that harbor IMBH. One example is the compact starburst galaxy Henize 2-10 \citep{reines16}. A signature of early episodes of positive feedback would be enhanced star formation. Some of the lowest mass  dwarfs have enhanced stellar mass fractions \citep{oman16}, for which there is no conventional  (i.e. stellar physics) explanation other than the possibility of systematic measurement error due to inclination.

\item
ULXs are  often found in  the outskirts of galaxies. This is also the domain of accreted ultrafaint dwarf galaxies. While some 
ULXs are almost certainly  neutron stars accreting at super-Eddington rates, some are considered as   likely IMBH candidates accreting from a close companion star \citep{bachetti16}. { \color{black}An IMBH has been reported in the inner MWG \citep{oka2016}
as well as in the massive globular cluster 47 Tuc \citep{kiziltan2017}.}

\end{enumerate}
\section {Predictions}

I now provide a few predictions that potentially provide a test of whether IMBH indeed inhabit dwarf galaxies.

\subsubsection{Kinematic imprints on stellar orbits}
AGN feedback may leave a unique signature on stellar orbits. Indeed SN feedback drives potential irregularities that are unveiled by stellar kinematics
\citep{elbadry16}.
 However the imprint of AGN feedback is more coherent, especially in the presence of  positive feedback
 \citep{dugan14}.
 Early AGN feedback in  nearby dwarf galaxies should  leave long-lived 
kinematic signatures on old stars that formed during the feedback phase.
Quenched jets or winds are similarly expected to perturb  velocities of newly forming stars. 
\textcolor{black}{
Low power jets (as well as outflows) from the AGN  are especially promising as they result from minor accretion events,  should greatly outnumber strong jets and are trapped in the interstellar medium for a longer time \citep{mukherjee2016}.} 

{\color{black}A simple estimate of the perturbed  stellar kinematics is as follows.  I assume the AGN outflow is described by a spherically symmetric   wind-blown interstellar bubble, the pressure-driven snowplow 
model of \cite{weaver1977}, and applied to triggered star formation by \cite {mccray1987}. For AGN triggering, the injected power is $\dot P=\eta L_{Edd}/c, $ where $\eta\sim 0.1\eta_{0.1}$ is the mechanical efficiency and $f_{Edd} L_{Edd}$ is the AGN luminosity relative to  Eddington of the accreting IMBH of mass $\rm M_{IMBH}=1000\rm M_\odot M_{IMBH,3}.$ The analytical solution for the shock velocity is $v_{sh} =(3\dot P/8\pi \rho_a\delta)^{1/2}r^{-1}$ for a cloud of density contrast $\delta =100\delta_{100}$ at distance $r$ from the IMBH. Assuming an isothermal profile for gas and dark matter, the cloud and stars formed therein receives an impulsive kick of velocity contrast relative to the dwarf circular velocity of
$$\sim 0.1 \eta_{0.1}^{1/2} ({f_{Edd} M_{IMBH,3}})^{1/2} M_{d,7}^{-2/3}\delta_{100}^{-1/2} z_{gf,10}^{-1} $$ 
for a dwarf of mass $10^7 M_{d,7}\rm M_\odot$ that formed at redshift $10 z_{gf,10}.$
This velocity perturbation will be long-lived, radially directed and potentially measurable as blue asymmetries in stellar absorption  line widths via IFU-type spectroscopy for compact dwarfs. A similar  effect  has been detected  in {\color{black} stacked} SDSS star-forming galaxies \citep{cicone2016}.
}
\textcolor{black}{
Dynamical detection via stellar kinematical signatures in  nearby compact dwarfs must await the next generation of  ELT-like telescopes \citep{volonteri2010}.}
\subsubsection{Tidal disruption events}
IMBH can tidally disrupt white dwarfs, unlike their SMBH counterparts.
The main difference with TDEs of main sequence stars is that a similar amount of mass is accreted on a much more rapid timescale. This leads to luminous flares, \textcolor{black}{
especially via tidal disruption of the hydrogen-rich envelopes often surrounding white dwarfs  and observed as nuclear transients on short time-scales \citep{lawsmith2017}}. Tidal decompression can trigger SNe {\color{black} with possible implications for acceleration of cosmic rays to ultrahigh energies{\color{black}, and in particular, an intermediate/heavy composition
for the observed UHECRs that is independent of specific acceleration models}
 \citep{batista2017}.} The debris of magnetized white dwarfs is subject to MRI-driven dynamo amplification of magnetic fields that produces radio jets \citep{shcherbakov12}

In a flux-limited survey of TDEs, white dwarf disruptions should account for $\sim 10\%$ of the events, and possibly more if IMBH are as numerous as our dwarf hypothesis implies. Jetted tidal disruption events 
 of main sequence stars  by binary  IMBH in dwarf galaxies, discussed  by \cite{fialkov16}, are likely to be outnumbered  by the more luminous  TDE signatures of white dwarfs \citep{krolik12}, which  also provide a prime mechanism for growing the IMBH \citep{baumgardt06}.

\noindent\subsubsection{Other possibilities}
I briefly review other promising predictions for exploring the presence of IMBH in dwarfs.


Gravitational microlensing against extended radio sources is a direct way of imaging IMBH. One proposal  requires  mapping strongly lensed radio-loud QSOs  at $10\mu $s  resolution  with a VLBI space observatory \citep {inoue03}. Another  approach  at much higher frequencies requires  the  next generation of submillimeter telescopes \citep{inoue13}.

Gravitational waves from IMBH mergers are a primary goal for LISA, expected to be launched in 2034 \citep{tinto16}. The dwarf galaxy ansatz augments the number of IMBH by of order $M_{SMBH}/M_{IMBH} \gtsim 100-1000,$ with potentially important implications for the predicted frequency of  \textcolor{black}{extreme-mass-ratio inspiral merger events that can probe the IMBH mass function in the range $10^4-10^7\rm M_\odot$ \citep{gair2010}.}


Tidal  stripping of dwarfs containing  relic IMBH in halos such as  M31 can leave morphologically distinct stellar debris signatures  that survive for hundreds of millions of years
 \citep{miki14}. High resolution simulations are required to study the signatures of wandering IMBH, \textcolor{black}{as might also be produced by infall of bulgeless disks \citep{kormendyho2013}.}


The source of reionization  in the early universe is unknown, because of the large uncertainty in escape fraction of ionizing photons from star-forming galaxies, \textcolor{black}{although dwarf galaxies are generally believed to be the dominant contributor given current optical depth measurements \citep{planck16}.  However  AGN are expected to have a high ionizing photon escape fraction, and SMBH alone, including their growth phase, are believed to  contribute
significantly to 
 the ionizing photon budget \citep{volonterignedin2009}. }If IMBH form early and are sufficiently numerous, as motivated by the present hypothesis, they can reionise the universe by production of hard photons. The mass in IMBH  expected  is  
of order the mass in low metallicity stars at $[Z]\ltsim  {-3},$ 
and  hence potentially   of comparable  efficiency to dwarf galaxies  in reionization of the universe.
The 21cm dark ages absorption signature against the CMB is  more pronounced  for soft photons than for hard photons with much longer mean free paths \citep{cohen16}.

\section{Summary}
SMBH scaling relations  have been extended to the IMBH regime. As previously noted, comparison of the $M_{SMBH}-\sigma$ and $M_{SMBH}-M_\ast$ relations shows that IMBH tend to lie below the extrapolated scaling relation
 \citep{baldassare15}, \textcolor{black}{and possibly also in the prototypical example  of  Henize 2-10 \citep{kormendyho2013}. }
\textcolor{black}{ The fact that $  M_\ast$ is effectively reduced in dwarfs as inferred from the scaling relation, admittedly with sparse data, strengthens the case for suppression of star formation in dwarfs,
and tends to confirm our hypothesis  that gas ejection may have played a greater role in dwarfs than in massive galaxies. }

This  might suggest that while SMBH may form contemporaneously with the old stellar population, IMBH could well have formed first. Confirmation of this trend  would strengthen the case for AGN feedback playing an important role in early dwarf evolution. {\color{black} The IMBH black hole occupation fraction inferred from x-ray observations of dwarfs is likely to be significant for any reasonable duty cycle: possible issues of black hole recoils will be considered elsewhere.}


\textcolor{black}{Dwarf galaxies continue to play a central role in cosmology.  For example, the most recent example is of 
the only known  repeating fast radio burst (FRB),  recently  localized by  VLBI/VLBA observations at sub-arc sec resolution \citep{chatterjee2017} to be in a dwarf galaxy host. This supports the inference that FRBs are extragalactic, one leading interpretation being that these highly luminous and short-duration events involve tidal disruption of a neutron star by a central IMBH \citep{romero2016}.  Localization of more FRBs, and especially detection of x-ray and optical emission line signatures, will lead to confirmation of the involvement of low luminosity AGN in dwarfs as a central hypothesis for understanding the inferred high frequency of FRBs.}

In summary, dwarf galaxies pose problems that are not easily resolved unless complicated and diverse  types of baryonic feedback are introduced. This has motivated many authors to invent dark matter modifications to explain such properties of dwarf galaxies as their frequency, structure and surface brightness. I argue that the rationale for such extreme measures, namely modifying fundamental physics,  is lacking  when plausible astrophysical explanations  remain to be explored. I provide one of these, which goes beyond the usual discussions of stellar feedback by appealing to a population of central IMBH that formed in the earliest gas-rich phase of dwarf galaxy evolution.  

The work of JS has been supported in part by  ERC Project No. 267117 (DARK)
hosted by the Pierre and  Marie Curie University-Paris VI, Sorbonne Universities and CEA-Saclay.
{\color{black}I thank P. Behroozi, M. Volonteri and A. Wagner  for illuminating discussions of these topics.}

%

\end{document}